\documentclass[aps,pra,reprint,superscriptaddress,amsmath,amssymb,floatfix]{revtex4-2}
\usepackage{graphicx}

\newcommand{\eps}{\varepsilon}

\begin{document}

\title{Single-anchor $\Delta$-learning with the cluster Gutzwiller approximation: predicting phase boundaries of strongly correlated bosons in optical lattices}

\author{Shengbiao Yang}
\thanks{ These authors have contributed equally to this work.}
\affiliation{School of Physics, Anhui University, Hefei 230601, China}

\author{Sheng Yue}
\thanks{ These authors have contributed equally to this work.}
\affiliation{School of Physics, Anhui University, Hefei 230601, China}
\affiliation{School of Physics, Hefei University of Technology, Hefei, Anhui 230009, China}

\author{Zhi Lin}
\email{zhilin18@ahu.edu.cn}
\affiliation{School of Physics, Anhui University, Hefei 230601, China}
\affiliation{State Key Laboratory of Surface Physics and Department of Physics, Fudan University, Shanghai 200433, China}

\begin{abstract}
The cluster Gutzwiller method is widely used to study strongly correlated bosonic systems in optical lattices because it explicitly treats quantum fluctuations within a cluster. However, its computational cost grows exponentially with the cluster size, making large-cluster phase diagrams computationally expensive. To reduce this cost, we recently introduced $\Delta$-learning into cluster Gutzwiller calculations and proposed multipoint regression $\Delta$-learning (MPR-$\Delta$), which learns the difference between small- and large-cluster results from four high-cost critical points of the target cluster and thereby predicts the complete target phase boundary. For sufficiently large target clusters, however, even computing one critical point on the phase boundary is already expensive, so obtaining four such points remains highly costly. A natural question then arises: can the complete target phase boundary be predicted with fewer high-cost critical points of the target cluster? To address this question, we propose single-anchor $\Delta$-learning (SA-$\Delta$), which requires only a single high-cost critical point of the target cluster to reconstruct the complete target phase boundary, thereby greatly reducing the computational cost. We test this framework on several classes of Bose--Hubbard models and find that, in the cases considered, it achieves lower reconstruction errors with fewer high-cost points than MPR-$\Delta$. The combination of SA-$\Delta$ with the cluster Gutzwiller approximation thus provides an efficient framework for studying phase transitions of strongly correlated bosonic systems in complex optical lattices, and may be extended to more complex quantum many-body systems.
\end{abstract}

\maketitle

\section{Introduction}\label{sec:intro}
Optical lattice systems loaded with ultracold atoms, whose dimensionality, geometry, and interaction strength can be independently tuned~\cite{Meinert2016,Clark2018}, have become an important experimental platform for testing conventional condensed-matter theory and for simulating intriguing quantum many-body physics that has no direct counterpart or is difficult to realize in conventional condensed-matter systems~\cite{Bloch2008,Lewenstein2012}. The experimental realization~\cite{Greiner2002} of the single-component Bose--Hubbard model~\cite{Fisher1989} has fully demonstrated the potential of strongly correlated bosonic systems in optical lattices for simulating such intriguing quantum many-body physics. When nearest-neighbor repulsion is introduced into this model, phases such as charge-density-wave (CDW) and supersolid (SS) also appear~\cite{Sengupta2005,Ohgoe2012}. Compared with the single-component Bose--Hubbard model, the two-component Bose--Hubbard model, realized by loading a two-component bosonic mixture into an optical lattice, exhibits a richer variety of phases, such as the two-component Mott insulator (2MI), two-component superfluid (2SF), paired superfluid (PSF)~\cite{Catani2008,Kuklov2003,Kuklov2004,Zheng2024}, molecular SF and noninteger MI phases~\cite{lin2020}, and two bosonic four-body superfluid phases: the paired super-counter-fluid (PSCF) and the symbiotic super-counter-fluid (SSCF)~\cite{PSCF2025}.

To study these strongly correlated bosonic systems, a commonly used method is the Gutzwiller mean-field approach and its cluster extension~\cite{Luhmann2013}. In the cluster Gutzwiller method, exact diagonalization is performed within each cluster, while the clusters are coupled at the mean-field level, so that quantum fluctuations inside a cluster are treated exactly. The phase diagrams it yields are qualitatively consistent with those from rigorous numerical methods~\cite{Lin2026,Zhang2026} such as quantum Monte Carlo (QMC)~\cite{Batrouni1992,Capogrosso2008} and density matrix renormalization group~\cite{Kuehner1998,Piraud2015}, while its computational cost is much lower than those of these methods. However, its accuracy is limited by the cluster size. Single-site calculations are extremely cheap but neglect inter-site correlations, whereas obtaining more accurate results requires larger clusters. Because the dimension of the Hilbert space to be diagonalized grows exponentially with the cluster size, phase diagrams for large clusters become computationally expensive.

$\Delta$-learning offers a viable way to reduce this computational cost: by learning the difference between the results of two computational methods of different accuracy, it predicts the complete high-accuracy result from only a few high-accuracy samples, and it has been widely applied in quantum chemistry~\cite{Ramakrishnan2015,Dral2020}. Building on this idea, we recently introduced $\Delta$-learning into cluster Gutzwiller calculations~\cite{Lin2026} and proposed multipoint regression $\Delta$-learning (MPR-$\Delta$): a small cluster provides a low-cost baseline phase boundary, and $\epsilon$-support vector regression ($\epsilon$-SVR) learns the difference between the baseline and the target from four high-cost points of the target cluster. Compared with directly computing the target-cluster phase diagram, this method reduces the number of required high-cost points from several dozen to four. For sufficiently large target clusters, however, even computing a single critical point is already expensive, and obtaining four such points remains costly. A natural question then arises: can the complete target phase boundary be predicted with even fewer high-cost points---possibly with only one?

\begin{figure*}[!t]\centering
	\includegraphics[width=\textwidth]{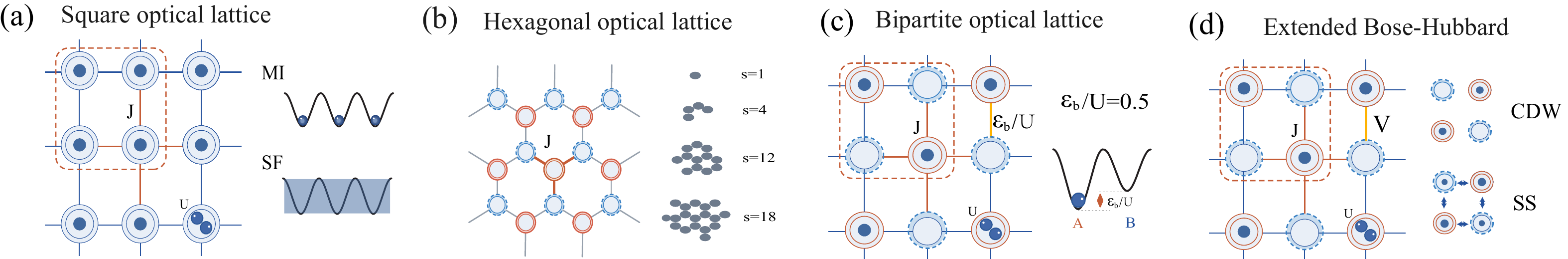}
	\caption{Lattice structures, clusters, and representative phases of the four classes of optical-lattice Bose--Hubbard models. In each panel the left side shows the lattice and the couplings; \(J\), \(U\), \(V\), and \(\eps_b\) denote nearest-neighbor tunneling, on-site repulsion, nearest-neighbor repulsion, and the sublattice energy offset, respectively. In (b)--(d), orange solid and blue dashed circles mark the \(A\) and \(B\) sublattices. (a) Square lattice (\(z=4\)): the upper-right and lower-right insets show the Mott insulating (MI) phase and the superfluid (SF) phase, respectively. (b) Hexagonal lattice (\(z=3\)): the right panel shows, from top to bottom, clusters with \(s=1\), \(4\), \(12\), and \(18\) sites, where \(s\) is the number of lattice sites in the cluster. (c) Bipartite optical lattice (staggered potential \(\eps_b/U=0.5\)): the right panel shows two wells of different depths, the difference in well depth being given by the staggered potential offset \(\eps_b\); the deep well (\(A\)) is singly occupied and the shallow well (\(B\)) is empty. (d) Extended Bose--Hubbard model (square lattice, \(z=4\); nearest-neighbor repulsion \(zV/U=1\)): the right panel shows the charge-density-wave (CDW) and supersolid (SS) phases, where the double-headed arrows in the SS sketch indicate phase coherence between sites. The red-orange dashed boxes in (a), (c), and (d) mark one representative \(2\times2\) cluster.}
	\label{fig:models}
\end{figure*}

Fortunately, we find that the answer is yes. In this work, we propose single-anchor $\Delta$-learning (SA-$\Delta$), which rests on the rank-one correction hypothesis. To formulate this hypothesis, let $J_{\rm m}(\mu)$ be the phase boundary of a reference small cluster, $J_{\rm b}(\mu)$ that of the baseline cluster, and $J_{\rm t}(\mu)$ that of the target cluster, and define the correction $D_1(\mu)=J_{\rm b}(\mu)-J_{\rm m}(\mu)$.
The rank-one correction hypothesis states that the correction of the target cluster relative to the baseline,
$\Delta(\mu)=J_{\rm t}(\mu)-J_{\rm b}(\mu)$, is approximately proportional to $D_1(\mu)$, i.e.,
$\Delta(\mu)\simeq \rho D_1(\mu)$,
where the proportionality coefficient $\rho$ is independent of the chemical potential $\mu$. In other words, increasing the cluster size mainly changes the amplitude of the correction, while the shape of its dependence on the chemical potential remains approximately unchanged. Under this hypothesis, the shape of the correction is given entirely by the difference of two directly computable low-cost curves, $D_1(\mu)$, without any high-cost calculation; the amplitude ratio $\rho$ is determined by a single high-cost calculation at a preselected anchor $\mu^\ast$. The entire target phase boundary is then reconstructed through the closed-form relation
$J_{\rm t}(\mu)\simeq J_{\rm b}(\mu)+\rho D_1(\mu)$.
This construction offers three advantages: the number of high-cost points is further reduced from four to one; the reconstruction is carried out by a closed-form expression, with no regression model to train and no hyperparameter search; and, as shown below, it even achieves higher reconstruction accuracy than MPR-$\Delta$. To assess the applicability of the method, we test it on four representative classes of bosonic lattice systems (Fig.~\ref{fig:models}): the Bose--Hubbard model on the square, hexagonal, and bipartite optical lattices, and the extended Bose--Hubbard model with nearest-neighbor repulsion. The results show that SA-$\Delta$ reconstructs the complete phase boundary with high accuracy in all four classes, and that in all cases where comparison with MPR-$\Delta$ is possible, its errors are lower than those of MPR-$\Delta$.

\begin{figure*}[!t]\centering
	\includegraphics[width=\textwidth]{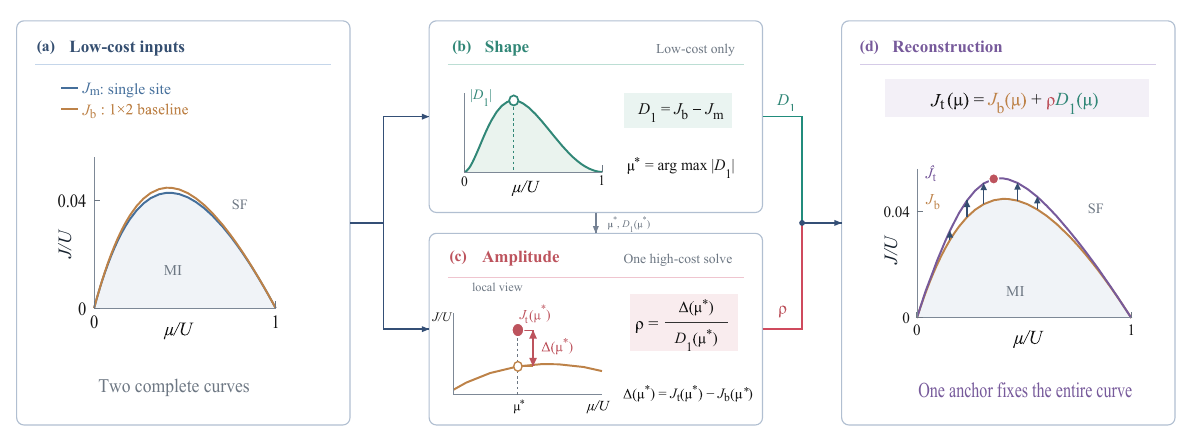}
	\caption{Workflow of SA-$\Delta$, illustrated for the square lattice. (a) Low-cost inputs: two complete low-cost curves, $J_{\rm m}$ (single site, blue) and $J_{\rm b}$ ($1\times2$ baseline, orange), both directly computable; the shaded region is the Mott insulating (MI) phase and the region outside it is the superfluid (SF) phase. (b) Shape (low-cost information only): the shape $D_1=J_{\rm b}-J_{\rm m}$ is given by the difference of the two low-cost curves, and its absolute value $|D_1|$ reaches its maximum at $\mu^\ast=\arg\max|D_1|$, marked by the green dot and the vertical dashed line. (c) Amplitude (one high-cost calculation): a local view around the anchor $\mu^\ast$, where the high-cost value $J_{\rm t}(\mu^\ast)$ (red dot) and its difference from the baseline, $\Delta(\mu^\ast)=J_{\rm t}(\mu^\ast)-J_{\rm b}(\mu^\ast)$, determine the amplitude ratio $\rho=\Delta(\mu^\ast)/D_1(\mu^\ast)$. (d) Reconstruction: the closed-form expression $J_{\rm t}(\mu)\simeq J_{\rm b}(\mu)+\rho D_1(\mu)$ corrects the baseline (blue arrows) to give the entire target phase boundary (purple solid line); the red dot is the anchor used, and one anchor fixes the entire curve.}
	\label{fig:flow}
\end{figure*}

The remainder of this paper is organized as follows. In Sec.~\ref{sec:method}, we briefly review the cluster Gutzwiller method and the finite-size scaling variable $\lambda$. In Sec.~\ref{sec:sadl}, we present single-anchor $\Delta$-learning, with the overall workflow illustrated in Fig.~\ref{fig:flow}. Section~\ref{sec:results} reports the results for the four classes of models. In Sec.~\ref{sec:compare}, we compare our method, within a unified framework, with the full cluster Gutzwiller calculation, the MPR-$\Delta$ of Ref.~\cite{Lin2026}, and external benchmarks such as QMC. Finally, Sec.~\ref{sec:conclusion} summarizes our findings.

\section{Cluster Gutzwiller method and the scaling variable}\label{sec:method}
Before introducing the construction of SA-\(\Delta\), we first review two preliminary ingredients: the cluster Gutzwiller method and the finite-size scaling variable \(\lambda\). The former is used to obtain, on clusters of different sizes, all the low-cost and high-cost phase boundaries required in this paper, while the latter characterizes the cluster size.

We first review the cluster Gutzwiller method, which is used to study strongly correlated bosonic systems in optical lattices. For the simplest single-component case, the Hamiltonian of the system can be written in the standard Bose--Hubbard form:
\begin{equation}
	\begin{split}
		\hat H_{\rm BH}=&-J\sum_{\langle ij\rangle}\bigl(\hat b_i^\dagger \hat b_j+{\rm h.c.}\bigr)\\
		&+\frac{U}{2}\sum_i \hat n_i(\hat n_i-1)-\mu\sum_i\hat n_i,
	\end{split}
	\label{eq:BH}
\end{equation}
where \(J\) is the nearest-neighbor tunneling strength, \(U\) the on-site repulsion strength, and \(\mu\) the chemical potential. Near integer filling, the ground state of the system is determined by the competition between tunneling and on-site repulsion: below a critical value of \(J/U\) the system is in the Mott insulating (MI) phase, while above it the system is in the superfluid (SF) phase, and the phase boundary is formed by the critical ratio \((J/U)_c(\mu)\) at each chemical potential \(\mu\). In this paper, we take \(U\) as the unit of energy, i.e., \(U=1\), and denote \((J/U)_c(\mu)\) by \(J_c(\mu)\). The \(J_c(\mu)\) curves obtained on small and large clusters are referred to as the low-cost and high-cost phase boundaries, respectively, because small clusters are cheap to compute while large clusters are expensive.

When solving for the \(J_c(\mu)\) curves of clusters of different sizes, the least expensive is the single-site Gutzwiller method, i.e., the limiting case of cluster size \(s=1\). It assumes that the wave function of the system factorizes into a product of on-site wave functions; the local state of site \(i\) is expanded in the Fock basis \(\{|n\rangle\}\) as \(|i\rangle=\sum_n c_n|n\rangle\), and the coefficients \(c_n\) are determined by self-consistent diagonalization. Under this assumption, the tunneling term is decoupled at the mean-field level, \(\hat b_i^\dagger\hat b_j\approx\hat b_i^\dagger\langle\hat b_j\rangle+\langle\hat b_i^\dagger\rangle\hat b_j-\langle\hat b_i^\dagger\rangle\langle\hat b_j\rangle\), so that the influence of the remaining sites on site \(i\) is carried entirely by the mean field \(\langle\hat b_j\rangle\), and the effective Hamiltonian of site \(i\) thus reads
\begin{equation}
	\hat H_i^{\rm eff}=\frac{U}{2}\hat n_i(\hat n_i-1)-\mu\hat n_i
	-J\Bigl(\hat b_i^\dagger\sum_{\langle j\rangle}\langle\hat b_j\rangle
	+\hat b_i\sum_{\langle j\rangle}\langle\hat b_j^\dagger\rangle\Bigr).
	\label{eq:gw}
\end{equation}
Where \(\langle\hat b_j\rangle=\sum_n c_n^*c_{n+1}\sqrt{n+1}\). Diagonalizing Eq.~\eqref{eq:gw} in the Fock basis yields the updated coefficients \(c_n\); these are then used to compute the order parameter \(\langle\hat b\rangle\), which in turn updates the effective Hamiltonian, and the cycle is repeated until self-consistency is reached.
Once self-consistency is reached, the ground-state wave function and the order parameter \(\langle\hat b\rangle\) are obtained at that parameter point \((J,U,\mu)\). A nonzero order parameter corresponds to the superfluid phase, whereas a vanishing order parameter yields the critical tunneling strength \(J_c(\mu)\) at that \(\mu\). This framework treats the on-site interaction and the number fluctuations exactly within a single site, but completely neglects the inter-site quantum correlations, and is therefore equivalent to single-site mean field.

To account for the inter-site quantum correlations neglected by the single-site approximation, the cluster Gutzwiller method~\cite{Luhmann2013} was subsequently developed. This method enlarges the basic unit of the product-state assumption from a single site to a cluster \(S\) containing \(s\) sites, and the cluster wave function can be expanded in the multisite Fock basis as \(|S\rangle=\sum_N C_N|N\rangle\). The internal degrees of freedom of the cluster are then retained completely, and the cluster Hamiltonian is diagonalized exactly in this Fock basis; only the bonds crossing the cluster boundary remain decoupled through the mean field \(\langle\hat b\rangle\), the number of mean-field couplings of a boundary site being determined by the number of its bonds that extend beyond the cluster. Following an iterative self-consistency procedure analogous to that of the single-site scheme, the quantum fluctuations and short-range correlations inside the cluster are treated exactly. The larger the cluster, the higher the fraction of bonds treated exactly, and hence the more accurate the resulting phase boundary; the price is that the dimension of the Hilbert space to be diagonalized grows exponentially with \(s\), and the computational cost of large-cluster phase boundaries therefore rises steeply.

We next introduce the finite-size scaling variable $\lambda$, which quantifies the finite-size effects arising from the cluster Gutzwiller calculation, and is defined as
\begin{equation}
	\lambda=\frac{B_S}{B_S+B_{\partial S}}.
	\label{eq:lambda}
\end{equation}
Where $B_S$ is the number of bonds inside the cluster $S$ and $B_{\partial S}$ the number of bonds crossing its boundary. Thus, $\lambda$ is the ratio of the number of bonds inside the cluster to the number of all bonds connected to the cluster. $\lambda=0$ corresponds to the single-site approximation and $\lambda\to1$ to the thermodynamic limit. Consequently, $\lambda$ provides both a common label for clusters of different sizes and a scaling variable for extrapolating the phase boundary to the thermodynamic limit.

We emphasize that, in the numerical implementation, the Fock-space truncation, the self-consistency tolerance, and the critical-point search procedure adopted in all single-site and cluster calculations of this paper follow those of Ref.~\cite{Lin2026}, so as to ensure that our results are directly comparable with theirs under the same numerical settings.

\section{Single-anchor \(\Delta\)-learning (SA-\(\Delta\))}\label{sec:sadl}
Building on the above solution framework and the scaling variable, this section presents the complete construction of SA-$\Delta$. Specifically, for a given target cluster, the method takes two directly computable low-cost curves as inputs and requires only one additional high-cost critical point on that cluster to reconstruct its phase boundary.

The starting point of this construction is the general form of $\Delta$-learning~\cite{Ramakrishnan2015,Dral2020}. In this form, the difference $\Delta_{\rm b}^{\rm t}$ between the target and the baseline is first learned from a few high-cost samples and then added to the low-cost baseline value $y_{\rm b}$, giving the target prediction $y_{\rm t}=y_{\rm b}+\Delta_{\rm b}^{\rm t}$. For the phase-boundary problem of interest in this paper, the low-cost prediction is the baseline phase boundary $J_{\rm b}(\mu)$ of a small cluster, the target prediction is the phase boundary $J_{\rm t}(\mu)$ of a large cluster, and the correction to be determined is the correction of the target cluster relative to the baseline, $\Delta(\mu)=J_{\rm t}(\mu)-J_{\rm b}(\mu)$. 

However, the definition of the correction $\Delta(\mu)$ involves the unknown target phase boundary $J_{\rm t}(\mu)$, so that $\Delta(\mu)$ cannot be obtained directly unless the large cluster is computed over the whole range; therefore its shape as a function of the chemical potential can only be inferred from low-cost curves. To this end, in addition to the baseline phase boundary $J_{\rm b}(\mu)$, we introduce a reference phase boundary $J_{\rm m}(\mu)$ given by a smaller cluster (e.g., a single site). The difference between the two is defined as the shape function:
\begin{equation}
	D_1(\mu)\equiv J_{\rm b}(\mu)-J_{\rm m}(\mu).
	\label{eq:D1}
\end{equation}
The reason for calling $D_1(\mu)$ a shape function is that it gives the functional form of the cluster-size correction as a function of the chemical potential $\mu$. More importantly, $D_1(\mu)$ depends only on the two low-cost curves of small clusters, and its computation requires no high-cost samples of the target cluster at all.

Whether the above construction is valid depends crucially on whether $D_1(\mu)$ reflects the true shape of the target correction $\Delta(\mu)$ as a function of the chemical potential. To test this point, we carry out systematic numerical calculations for clusters of different sizes (see the reconstruction results in Sec.~\ref{sec:results}). The calculations show that, as the cluster size increases, the shape of the phase-boundary correction as a function of the chemical potential $\mu$ remains approximately unchanged, and what changes is mainly its overall amplitude. Based on this fact, we approximate the correction of the target cluster as the product of the shape function $D_1(\mu)$ and an amplitude ratio $\rho$:
\begin{equation}
	\Delta(\mu)\simeq\rho D_1(\mu),\qquad
	\rho\equiv\frac{\Delta(\mu^\ast)}{D_1(\mu^\ast)}.
	\label{eq:rho}
\end{equation}
Where $\rho$ is an amplitude ratio that depends only on the cluster size, and $\mu^\ast$ is a preselected calibration anchor (whose choice is described below). We refer to this approximation as the rank-one correction hypothesis. If the phase-boundary correction is viewed as a function of the chemical potential and the cluster size, the hypothesis approximates it as the outer product of a shape function that depends only on the chemical potential and an amplitude factor that depends only on the cluster size, i.e., a rank-one decomposition. Moreover, the scaling variable $\lambda$ introduced above provides a physical motivation for this hypothesis: as $\lambda\to1$, the overall amplitude of the finite-size correction tends to zero while the shape of the correction itself always remains well defined; assuming that the shape does not change with the cluster size is therefore the natural zeroth-order approximation in a finite-size expansion. Thus, under the rank-one correction hypothesis, the only unknown in the target correction $\Delta(\mu)$ is the scalar amplitude ratio $\rho$; a single high-cost calculation on the target cluster suffices to determine it.

Once the amplitude ratio $\rho$ is determined, the entire target phase boundary can be reconstructed directly through the following closed-form expression:
\begin{equation}
	J_{\rm t}(\mu)\simeq J_{\rm b}(\mu)+\rho D_1(\mu).
	\label{eq:main}
\end{equation}
Where the amplitude ratio $\rho$ is calibrated from one high-cost calculation at the anchor $\mu^\ast$ according to Eq.~\eqref{eq:rho}, and the anchor is placed at the chemical potential where $|D_1(\mu)|$ is maximal:
\begin{equation}
	\mu^\ast=\arg\max_\mu|D_1(\mu)|.
	\label{eq:anchor}
\end{equation}
Choosing the anchor in this way has two advantages: on the one hand, the maximum of $|D_1(\mu)|$ corresponds to the region where the two low-cost curves of small clusters differ most, so that calibrating there reduces the influence of the error in the high-cost calculation on $\rho$; on the other hand, this also avoids the region where $D_1(\mu)$ is close to zero (e.g., the ends of a Mott lobe), and thus avoids the numerical ill-conditioning caused by a too-small denominator when computing $\rho=\Delta(\mu^\ast)/D_1(\mu^\ast)$.
Next, Fig.~\ref{fig:flow} illustrates the complete workflow of the above construction for the square lattice: the two low-cost curves $J_{\rm m}$ and $J_{\rm b}$ are computed first, and their difference gives $D_1(\mu)$ and determines the anchor $\mu^\ast$; one high-cost calculation is then performed on the target cluster at the anchor to calibrate $\rho$; finally, substituting into Eq.~\eqref{eq:main} reconstructs the entire target phase boundary. These steps correspond respectively to Figs.~\ref{fig:flow}(a)--(d).

To quantitatively evaluate the accuracy with which the above construction reconstructs the phase boundary, we adopt the mean absolute percentage error (MAPE) as a unified error measure:
\begin{equation}
	\mathrm{MAPE}=\frac{1}{N}\sum_{i=1}^{N}\left|\frac{y_i-\hat y_i}{y_i}\right|\times 100\%.
	\label{eq:mape}
\end{equation}
Where $y_i$ is the cluster Gutzwiller reference value, $\hat y_i$ the predicted value, and $N$ the number of scoring points. In the scoring, the first and last chemical-potential points of the target phase boundary are discarded, and only the points whose reference value is not lower than $5\%$ of the maximum value of the reference curve are retained, so as to reduce the interference of near-zero reference values with the error measure.

This completes the construction of SA-$\Delta$: reconstructing each target phase boundary requires only two low-cost curves and one high-cost calculation at the anchor; the reconstruction is carried out by the closed-form expression~\eqref{eq:main}, and its accuracy is measured uniformly by Eq.~\eqref{eq:mape}.

\section{Results for four classes of optical-lattice models}\label{sec:results}
To fully assess the applicability of SA-$\Delta$, we examine the accuracy with which it reconstructs phase boundaries on four representative classes of optical-lattice models, and extrapolate these boundaries to the thermodynamic limit by means of finite-size scaling. Of these four classes, the first two are homogeneous systems with equivalent lattice sites, namely the Bose--Hubbard model on the square and hexagonal optical lattices; the latter two are spatially inhomogeneous, but the inhomogeneity has different origins: one is the Bose--Hubbard model on the bipartite optical lattice with a staggered potential, whose inhomogeneity is introduced by an external periodic potential, and the other is the extended Bose--Hubbard model with nearest-neighbor repulsion, whose inhomogeneity arises from the spontaneous symmetry breaking induced by the repulsive interaction.

\subsection{Homogeneous systems}\label{sec:uniform}
We first consider the homogeneous systems with equivalent lattice sites, namely the square and hexagonal lattices. Both of these lattice systems, loaded with ultracold bosons, can be described by the Bose--Hubbard model, Eq.~\eqref{eq:BH}, but they differ in lattice geometry and in the nearest-neighbor connectivity, with coordination numbers \(z=4\) and \(z=3\), respectively (Figs.~\ref{fig:models}(a) and (b)). Both lattices serve to test the reconstruction accuracy as the target cluster size increases; moreover, because the square and hexagonal lattices differ in coordination number and connectivity, comparing them allows us to examine whether SA-$\Delta$ applies to different lattice geometries. Because no QMC benchmark is available for the hexagonal lattice, the finite-size extrapolation is performed only on the square lattice.

\begin{figure}\centering
	\includegraphics[width=\columnwidth]{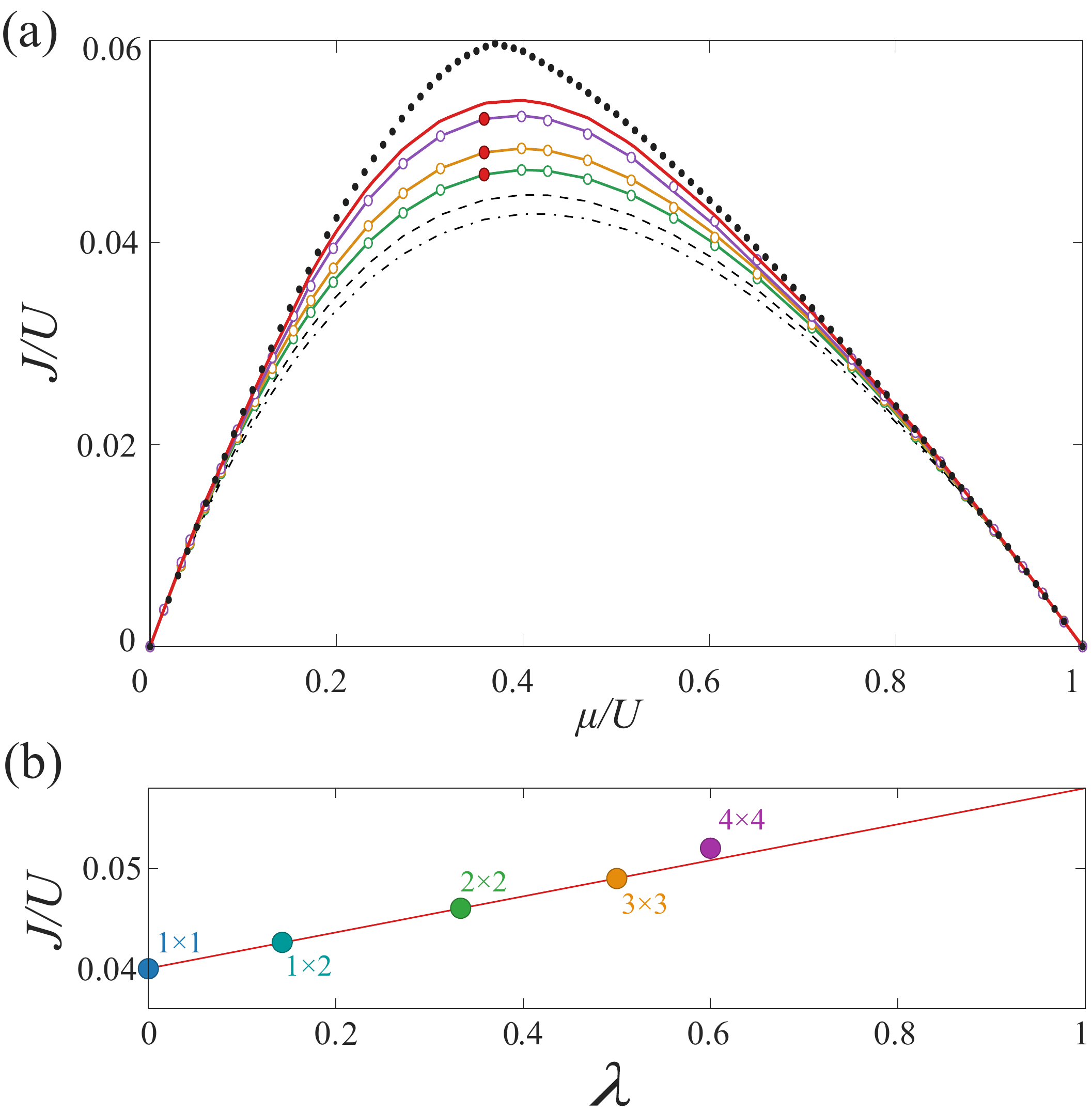}
	\caption{SA-$\Delta$ reconstruction on the square lattice. (a) The green, gold, and purple solid lines represent the phase boundaries predicted by SA-$\Delta$ for the $2\times2$, $3\times3$, and $4\times4$ clusters, respectively, and the open circles of the same colors indicate the results calculated directly by the cluster Gutzwiller method for the same clusters. The three red filled circles denote the anchor points of the three clusters; all three are located at the same chemical potential, $\mu^\ast/U=0.358$. The black dashed and dash-dotted lines represent the low-cost curves of the $1\times2$ cluster and the single site, respectively. The black dots are the QMC results, taken from Ref.~\cite{Capogrosso2008}. The red solid line is the thermodynamic-limit phase boundary
	obtained by finite-size scaling extrapolation. (b) Finite-size scaling of the critical value
	$J_c(\mu^\ast)$ at the anchor. The filled circles are colored, from the smallest to the largest cluster, blue, cyan, green, gold, and purple, and the cluster size is labeled next to each point; the red solid line is the linear extrapolation to $\lambda=1$. The scaling variable $\lambda$ equals zero for a single-site cluster and approaches one in the thermodynamic limit.}
	\label{fig:square-rungs}
\end{figure}

On the square lattice, we use the phase boundaries of the $1\times1$ cluster (used to obtain the $J_{\rm m}$ curve) and of the $1\times2$ cluster (used to obtain the $J_{\rm b}$ curve) as the low-cost inputs, perform single-anchor reconstruction for target clusters of the three sizes $2\times2$, $3\times3$, and $4\times4$, and set the anchor to $\mu^\ast/U=0.358$ for all three. All three prediction curves are in excellent agreement with the corresponding cluster Gutzwiller reference results on both sides of the Mott lobe and near its tip [Fig.~\ref{fig:square-rungs}(a)], with MAPE values of \(0.308\%\), \(1.045\%\), and \(1.840\%\), respectively. Therefore, within the range of cluster sizes examined here, a single anchor not only fixes the local phase boundary but also yields an accurate account of the entire boundary, although the reconstruction error increases with the target cluster size.

\begin{figure}[!t]\centering
	\includegraphics[width=\columnwidth]{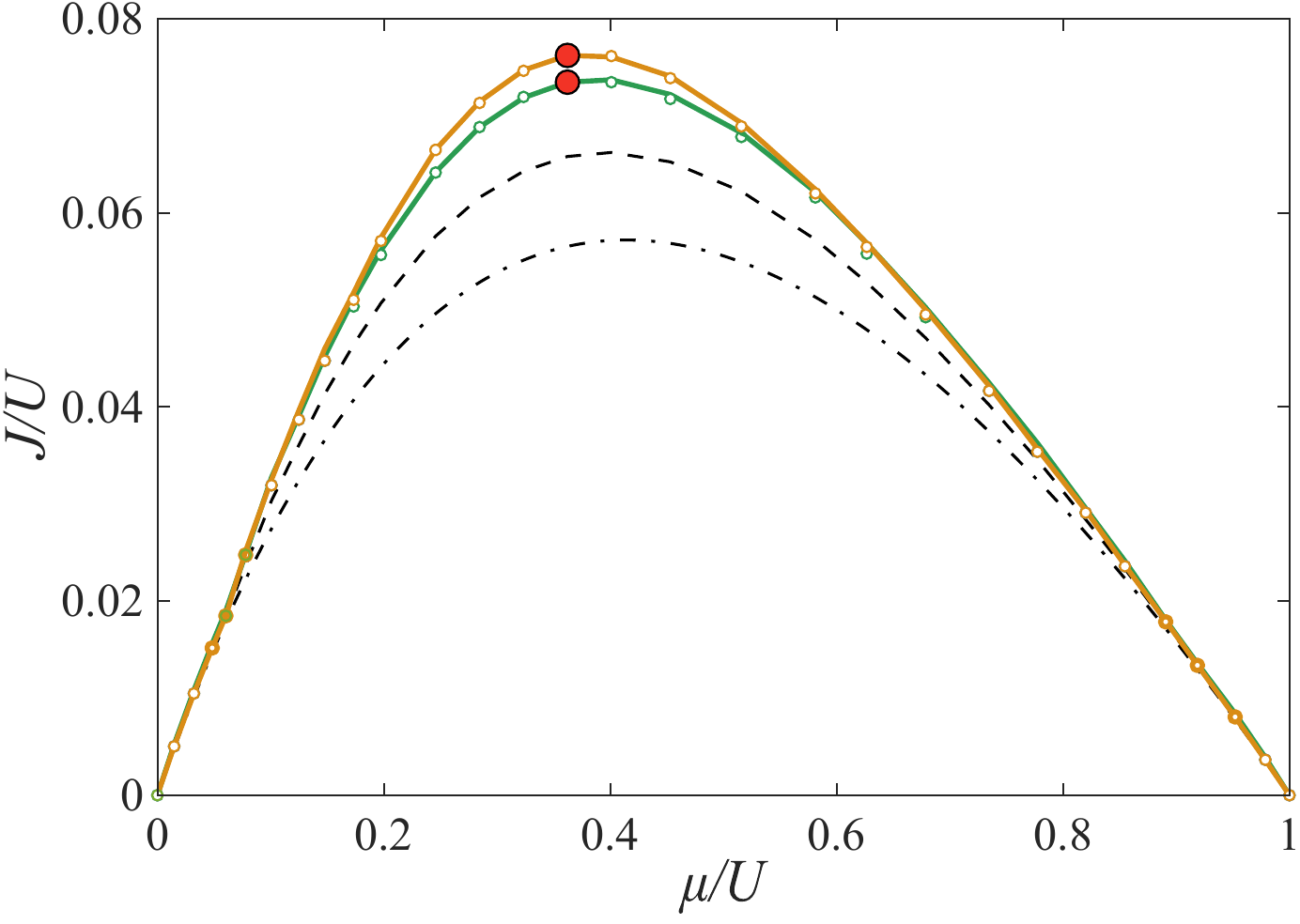}
	\caption{SA-$\Delta$ reconstruction on the hexagonal lattice. The green and gold solid lines
		represent the phase boundaries predicted by SA-$\Delta$ for the $s=12$ ($3\times4$) and $s=18$ ($3\times6$) clusters, respectively, and the open circles of the same colors indicate the results calculated directly by the cluster Gutzwiller method for the same clusters. The two red filled circles denote the anchor points of the two clusters; both are located at the same chemical potential, $\mu^\ast/U=0.359$. The black dashed and dash-dotted lines represent the low-cost curves of the $s=4$ ($2\times2$) and $s=1$ (single site) clusters, respectively.}
	\label{fig:hex-rungs}
\end{figure}

Building on the results for finite clusters, we further estimate the thermodynamic-limit phase boundary. Specifically, we extrapolate the phase-boundary values $J_c(\mu^\ast)/U$ of clusters of different sizes at the same anchor linearly in $\lambda$, obtaining the estimate at $\lambda=1$ [Fig.~\ref{fig:square-rungs}(b)]. With the low-cost shape $D_1$ kept unchanged, we then replace the high-cost value at the anchor with this extrapolated value and recalibrate $\rho$ according to Eq.~\eqref{eq:rho}; the resulting boundary agrees well with the QMC benchmark~\cite{Capogrosso2008} on both sides of the Mott lobe, but a visible deviation remains near the tip.

For the hexagonal lattice~\cite{Soltan2011,Soltan2012}, changing the lattice geometry requires the two low-cost curves and their difference $D_1$ to be recomputed, whereas the reconstruction procedure itself remains unchanged. We use the phase boundaries of the $s=1$ cluster (single site, used to obtain the $J_{\rm m}$ curve) and the $s=4$ cluster ($2\times2$, used to obtain the $J_{\rm b}$ curve) as the low-cost inputs, perform single-anchor reconstruction for target clusters of the two sizes $s=12$ ($3\times4$) and $s=18$ ($3\times6$), and set the anchor to $\mu^\ast/U=0.359$ for both. Both prediction curves are in excellent agreement with the corresponding cluster Gutzwiller reference results [Fig.~\ref{fig:hex-rungs}], with MAPE values of \(1.620\%\) and \(2.858\%\), respectively. As in the square-lattice case, the MAPE also increases with the target cluster size. Taken together, the results for the square and hexagonal lattices show that SA-$\Delta$ reconstructs the complete phase boundary accurately from a single anchor on both homogeneous lattices, which differ in coordination number and connectivity, indicating that its validity is not limited to a particular lattice geometry.

\subsection{Inhomogeneous systems}\label{sec:inhomogeneous}
The results for homogeneous lattices do not yet establish whether SA-$\Delta$ can handle sublattice inequivalence or phase boundaries of a different type. To this end, we further examine two spatially inhomogeneous systems, whose inhomogeneity has different origins: one is the bipartite optical lattice with a staggered potential, in which the staggered potential makes the two sublattices inequivalent; the other is the extended Bose--Hubbard model with nearest-neighbor repulsion, in which the nearest-neighbor repulsion gives rise to CDW and SS phases. The former is used to test how the spatial inhomogeneity caused by sublattice inequivalence affects the reconstruction, whereas the latter is used to test whether the method can handle the more complex phase boundaries induced by nearest-neighbor repulsion, such as the SS--SF phase boundary.

\begin{figure}[!t]\centering
	\includegraphics[width=\columnwidth]{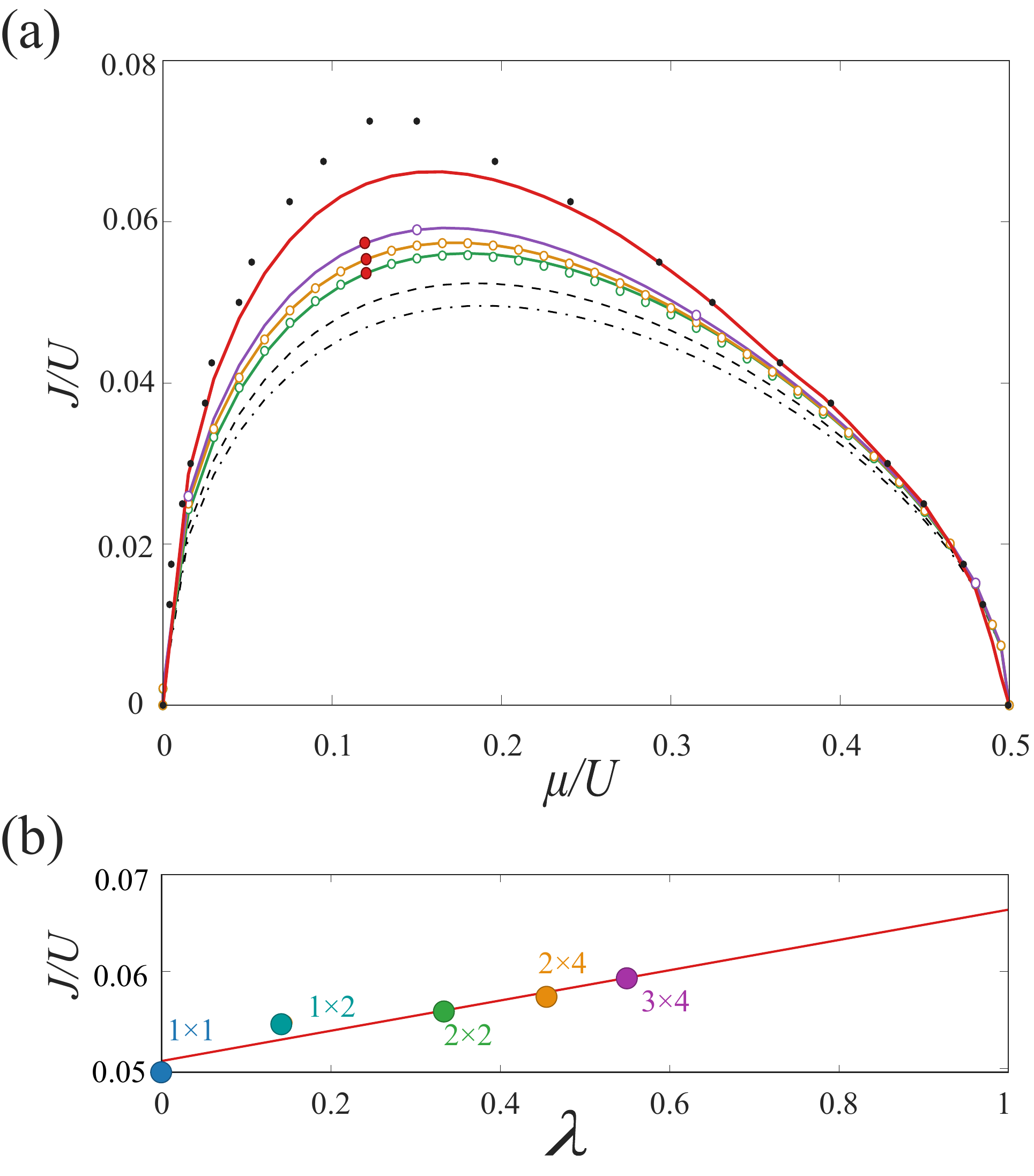}
	\caption{SA-$\Delta$ reconstruction on the bipartite optical lattice. (a) The green, gold, and purple solid lines represent the phase boundaries predicted by SA-$\Delta$ for the $2\times2$, $2\times4$, and $3\times4$ clusters, respectively, and the open circles of the same colors indicate the results calculated directly by the cluster Gutzwiller method for the same clusters. The three red filled circles denote the anchor points of the three clusters; all three are located at the same chemical potential, $\mu^\ast/U=0.120$. The black dots are the QMC results given in Ref.~\cite{Wang2013}. The red solid line is the thermodynamic-limit phase boundary obtained by finite-size scaling extrapolation. The black dashed and dash-dotted lines represent the low-cost curves of the $1\times2$ and $1\times1$ clusters, respectively. (b) Finite-size scaling of the critical value $J_c(\mu^\ast)$ at the anchor. The filled circles are colored, from the smallest to the largest cluster, blue, cyan, green, gold, and purple, and the cluster size is labeled next to each point; the red solid line is the linear extrapolation to $\lambda=1$.}
	\label{fig:bip-threeway}
\end{figure}

The bipartite optical lattice model~\cite{DiLiberto2014} introduces an energy offset between the two sublattices $A$ and $B$ on top of the homogeneous square lattice, making them inequivalent. The Hamiltonian of this model reads
\begin{equation}
	\begin{split}
		\hat H = & -J\sum_{i\in A,\,j\in B}\bigl(\hat b_i^\dagger \hat b_j + {\rm h.c.}\bigr)
		+ \frac U2\sum_{i\in A\cup B} \hat n_i(\hat n_i-1) \\
		& -\mu\sum_{i\in A}\hat n_i - (\mu-\eps_b)\sum_{i\in B}\hat n_i.
	\end{split}
	\label{eq:bip-H}
\end{equation}
Where the tunneling term includes only nearest-neighbor $A$--$B$ site pairs, $\eps_b$ is the staggered potential offset between the two sublattices, and, accordingly, the local chemical potentials of the $A$ and $B$ sublattices are $\mu$ and $\mu-\eps_b$, respectively. This Hamiltonian corresponds to the bipartite optical lattice model studied in Ref.~\cite{Lin2026}, except that the chemical potential is shifted globally here so that the first Mott lobe lies in a positive chemical-potential interval. With $\eps_b/U=0.5$, in the zero-tunneling limit ($J=0$), the system is in the CDW state for $0<\mu/U<0.5$, i.e., each $A$ site is singly occupied and each $B$ site is empty.

Figure~\ref{fig:bip-threeway}(a) shows the low-cost input curves obtained from the phase boundaries of the $1\times1$ and $1\times2$ clusters, whose difference gives the shape function $D_1(\mu)$ required for the reconstruction. On this basis, we perform single-anchor reconstruction for three target clusters ($2\times2$, $2\times4$, and $3\times4$) and compare the reconstructed curves with the corresponding cluster Gutzwiller reference values. For all three target clusters, the anchor is determined from Eq.~\eqref{eq:anchor} to be $\mu^\ast/U=0.120$. At this anchor, the predicted boundaries of the $2\times2$ and $2\times4$ clusters are both in excellent agreement with the reference results, with MAPE values of $0.014\%$ and $0.261\%$, respectively. For the $3\times4$ cluster, the complete curve is likewise reconstructed with this anchor, but only five reference points have been computed thus far because of the high cost of the target-cluster calculation; under the scoring rule described above, four of these points enter the scoring, giving a MAPE of $0.597\%$. This indicates that the reconstruction is reasonably accurate at the computed points, but it cannot yet be validated over the full range of the reference curve. Because these four reference points also serve as the training points of MPR-$\Delta$, this case is not included in the accuracy comparison between the two methods.

For the thermodynamic limit, we extrapolate the critical values of the clusters at $\mu^\ast/U=0.120$ linearly in $\lambda$, and then calibrate the boundary with the extrapolated value [Fig.~\ref{fig:bip-threeway}(b)]. The resulting thermodynamic-limit boundary is in quantitative agreement with the QMC benchmark~\cite{Wang2013} on both sides of the Mott lobe, with a visible deviation only near the tip.

We next consider the extended Bose--Hubbard model~\cite{Iskin2009,LinYang2019}, whose structure is shown in Fig.~\ref{fig:models}(d): a nearest-neighbor repulsion \(V\) is added on top of the homogeneous square lattice, and the Hamiltonian reads
\begin{equation}
	\begin{split}
		\hat H=&-J\sum_{\langle ij\rangle}\bigl(\hat b_i^\dagger \hat b_j+{\rm h.c.}\bigr)+\frac U2\sum_i \hat n_i(\hat n_i-1)\\
		&+V\sum_{\langle ij\rangle}\hat n_i\hat n_j-\mu\sum_i \hat n_i .
	\end{split}
	\label{eq:ebh-H}
\end{equation}
Where $z=4$ is the coordination number and we take $zV/U=1$. For these parameters, the first lobe is a CDW$(1,0)$ (with occupation numbers 1 and 0 on the two sublattices, respectively), and an SS phase appears between the CDW and SF phases. For this model, we examine three types of phase boundaries separately: CDW--SF, CDW--SS, and SS--SF.

\begin{figure*}[!t]\centering
	\includegraphics[width=\textwidth]{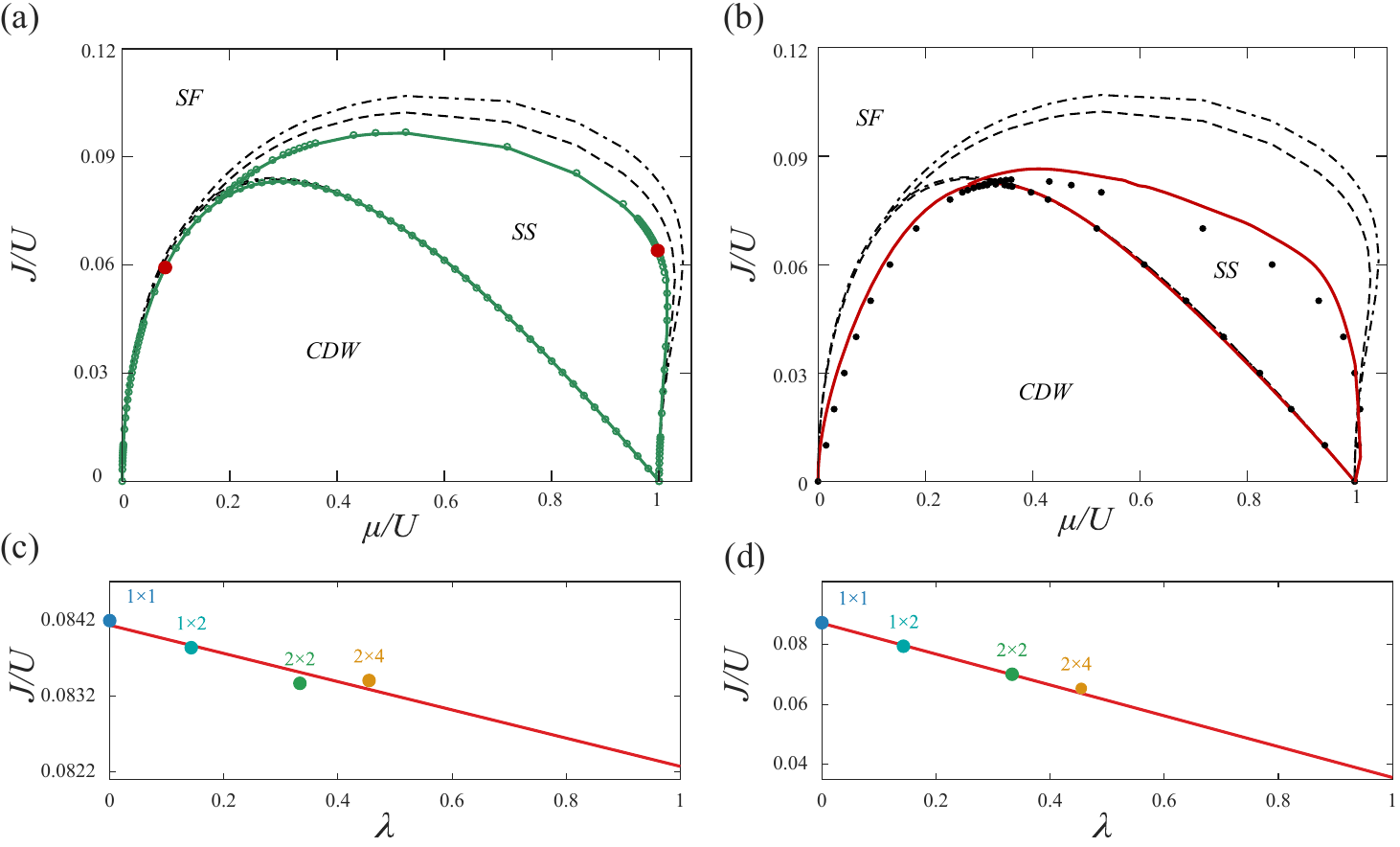}
	\caption{SA-$\Delta$ reconstruction in the extended Bose--Hubbard model ($zV/U=1$). (a) The green solid lines represent the CDW and SS--SF boundaries predicted by SA-$\Delta$ for the $2\times2$ cluster, and the open circles of the same color indicate the results calculated directly by the cluster Gutzwiller method for the same cluster. The two red filled circles denote the anchor points of the two boundaries, located at $\mu^\ast/U=0.080$ and \(0.975\), respectively. The black dashed and dash-dotted lines represent the low-cost curves of the $1\times2$ and $1\times1$ clusters, respectively. (b) The red solid lines are the thermodynamic-limit phase boundaries obtained by extrapolation to $\lambda\to1$. The black dots are the QMC results, taken from Ref.~\cite{Ohgoe2012}. The low-cost curves are the same as in (a). (c) Finite-size scaling of the CDW boundary at the anchor. The filled circles are the critical values $J_c(\mu^\ast)$ of the $1\times1$, $1\times2$, $2\times2$, and $2\times4$ clusters, and the red solid line is the linear extrapolation to $\lambda=1$. (d) Finite-size scaling of the SS--SF boundary at the anchor. The filled circles are the critical values of the same four clusters, and the red solid line is the
	linear extrapolation to $\lambda=1$.}
	\label{fig:ebh-rungs}
\end{figure*}

Unlike the bipartite optical lattice, the extended Bose--Hubbard model exhibits a complete CDW lobe in its phase diagram, and the type of phase transition on this lobe changes with the chemical potential: at low chemical potentials the boundary is CDW--SF, whereas beyond a certain critical value it becomes CDW--SS (Fig.~\ref{fig:ebh-rungs}). Therefore, for this system, whether a single calibration can cover the entire boundary to be reconstructed still requires further examination. Using the low-cost curves of the $1\times1$ and $1\times2$ clusters as inputs, we perform single-anchor reconstruction for the $2\times2$ target cluster, with the anchor at $\mu^\ast/U=0.080$, which lies on the CDW--SF boundary. As the chemical potential increases, the SS phase sets in and the boundary becomes CDW--SS. The curve obtained from a single calibration covers both segments at once, with a MAPE of $0.231\%$ relative to the cluster Gutzwiller reference results [Fig.~\ref{fig:ebh-rungs}(a)], indicating that the same amplitude calibration applies equally to boundary segments of different physical nature on the same lobe. We also perform a separate single-anchor reconstruction for the upper SS--SF boundary, with the anchor at $\mu^\ast/U=0.975$ [the red filled circle on the right in Fig.~\ref{fig:ebh-rungs}(a)], which gives a MAPE of $0.113\%$ for the $2\times2$ cluster; the errors are summarized in Fig.~\ref{fig:cmp-acc}. We emphasize that the CDW boundary and the SS--SF boundary are two independent reconstruction tasks, for which the low-cost shape and the calibration anchor must be determined separately; the requirement of ``one high-cost point per boundary'' therefore applies to each boundary to be reconstructed, rather than to the entire multiphase phase diagram.

Figure~\ref{fig:ebh-rungs}(b) further shows the thermodynamic-limit CDW--SS and SS--SF phase boundaries obtained from the $\lambda\to1$ scaling extrapolation and compares them with the QMC reference results~\cite{Ohgoe2012}: the two are in excellent agreement in terms of their overall trends, with the CDW--SS boundary in quantitative agreement with the QMC data at the lobe tip and in the region to its right. Figures~\ref{fig:ebh-rungs}(c) and \ref{fig:ebh-rungs}(d), in turn, show the finite-size convergence of the critical values of the CDW lobe boundary and of the SS island boundary at their respective anchors as functions of the scaling variable $\lambda$, together with the corresponding linear extrapolations; extrapolating to $\lambda=1$ gives reasonable estimates of the thermodynamic limit. These results show that SA-$\Delta$ is equally effective for the extended Bose--Hubbard model with nearest-neighbor repulsion, and that the applicability of the rank-one correction hypothesis is not restricted to conventional Mott-type transitions but also covers transitions involving the CDW and SS ordered phases, including the three types of phase boundaries CDW--SF, CDW--SS, and SS--SF.

In summary, for all four classes of homogeneous and inhomogeneous optical-lattice models, SA-$\Delta$ reconstructs the complete phase boundary with high accuracy from a single high-cost anchor. For the square lattice, the bipartite optical lattice, and the extended Bose--Hubbard model, the thermodynamic-limit boundaries obtained in combination with the $\lambda$ scaling extrapolation agree well with the QMC benchmarks (no QMC benchmark is yet available for the hexagonal lattice). These results together support the applicability of the rank-one correction hypothesis in the cases examined. Moreover, the reconstruction error in each system increases with the target cluster size; for example, on the square lattice the MAPE grows from $0.308\%$ for the $2\times2$ cluster to $1.840\%$ for the $4\times4$ cluster. This trend arises because the correction amplitude itself becomes smaller for larger clusters, so that the same absolute deviation produces a larger relative error, and higher-order corrections to the shape begin to accumulate. Nevertheless, these deviations remain very small; if higher accuracy is required, the two low-cost curves can be taken from larger clusters instead. This trend does not affect the conclusions drawn for the target size range of this paper, while the good agreement between the thermodynamic-limit boundaries and the QMC benchmarks further indicates that, in the systems examined here, the approximate invariance of the correction shape holds all the way to the thermodynamic limit.

\section{Unified comparison with existing approaches}\label{sec:compare}
The above results show that a single high-cost anchor can yield a high-accuracy reconstruction of the phase boundary in the cases examined. What needs to be answered next is this: compared with the multipoint regression $\Delta$-learning method (MPR-$\Delta$), does this simplification bring advantages in accuracy and computational cost at the same time? To this end, we compare SA-$\Delta$ with the MPR-$\Delta$ of Ref.~\cite{Lin2026} in terms of both accuracy and computational cost under the same cluster Gutzwiller solution conditions. In this section, the cluster Gutzwiller results for the target clusters serve as the reference for the reconstruction error, while the computational cost is measured primarily by the number of high-cost points required on the target cluster; the external benchmarks of the previous section, such as QMC, serve mainly to test the $\lambda\to1$ extrapolation to the thermodynamic limit. The former measures the reconstruction accuracy on finite clusters, whereas the latter tests the reliability of the extrapolation; the two play different roles.

\subsection{Accuracy}\label{sec:cmp-acc}
The accuracy comparison uses the MAPE and the scoring-point screening rule defined in Sec.~\ref{sec:sadl}, excluding the high-cost points required by the construction (the anchor of SA-$\Delta$ and the training points of MPR-$\Delta$); all remaining reference values are used for the error evaluation. Two cases require special comment: the square-lattice $4\times4$ case has no MPR-$\Delta$ comparison results (Ref.~\cite{Lin2026} did not provide results for this size); the bipartite optical lattice $3\times4$ case has only four reference points entering the scoring, and these are exactly the four training points of MPR-$\Delta$, so that no scoring points independent of the training are available; it is therefore excluded from the accuracy comparison between the two methods, and only the reconstruction error of SA-$\Delta$ at these four reference points is reported. Therefore, eight of the ten SA-$\Delta$ cases can be compared in accuracy with MPR-$\Delta$.

\begin{figure}[!t]\centering
	\includegraphics[width=\columnwidth]{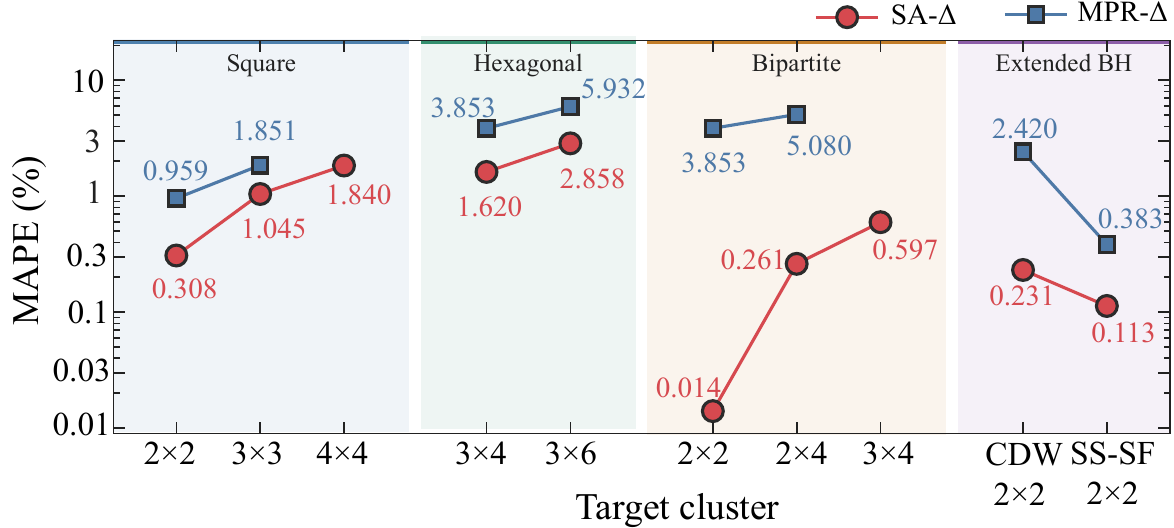}
	\caption{Comparison of the reconstruction accuracy for the four optical-lattice models. The
		horizontal axis represents the target cluster and the vertical axis the MAPE (\(\%\), logarithmic scale). The lightly shaded regions, from left to right, denote the square lattice, the hexagonal lattice, the bipartite optical lattice, and the extended Bose--Hubbard model; the last group corresponds to the CDW and SS--SF boundaries of the $2\times2$ cluster. The red circles and blue squares represent the results of SA-$\Delta$ (one anchor) and MPR-$\Delta$ (four training points)~\cite{Lin2026}, respectively; the connecting lines link only results obtained with the same method within the same model, and the numbers give the corresponding MAPE values. No MPR-$\Delta$ results are available for the square-lattice $4\times4$ and bipartite $3\times4$ cases; for the latter, only the MAPE over the four reference points is reported.}
	\label{fig:cmp-acc}
\end{figure}

In all eight comparable cases, SA-$\Delta$ obtains a lower MAPE with one high-cost point than MPR-$\Delta$ does with its multipoint regression [Fig.~\ref{fig:cmp-acc}]. The errors reported for all ten SA-$\Delta$ cases range from $0.014\%$ to $2.858\%$; the value for the bipartite optical lattice $3\times4$ case represents only the MAPE over the four reference points. For the SS--SF boundary discussed in Sec.~\ref{sec:results}, the MAPE values of SA-$\Delta$ and MPR-$\Delta$ are $0.113\%$ and $0.383\%$, respectively, indicating that this accuracy advantage also holds on this boundary.

It should be pointed out that the MPR-$\Delta$ errors used for comparison are obtained in two ways: some of them are recomputed from the prediction curves of Ref.~\cite{Lin2026} with the evaluation protocol of this paper, and the rest are obtained by rerunning that method and evaluating it with the same protocol. The above comparison shows that, in these cases, the use of the low-cost shape constraint not only reduces the number of high-cost points but also lowers the reconstruction error.

\begin{figure}[!t]\centering
	\includegraphics[width=\columnwidth]{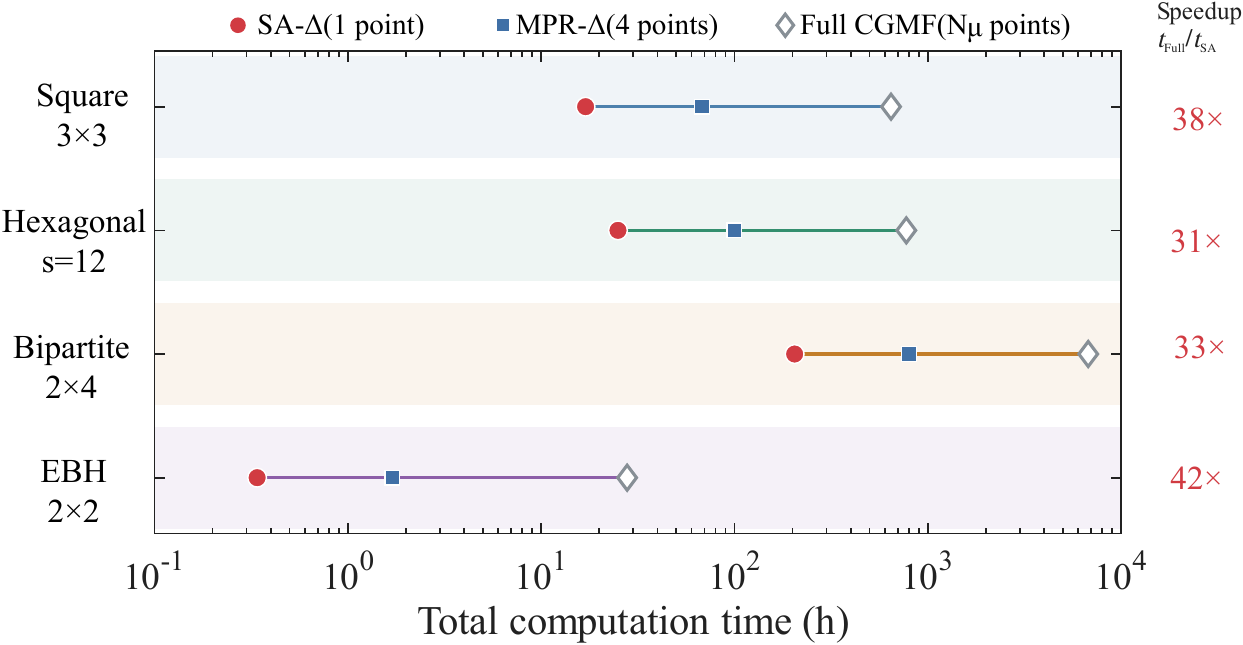}
	\caption{Comparison of the computational cost for the four target systems. The horizontal axis represents the total computation time required for one complete target phase boundary (logarithmic scale, in hours; the calculations were performed on a 10th-generation Intel Core i7). From top to bottom, the rows correspond to the square lattice $3\times3$, the hexagonal lattice $s=12$ ($3\times4$), the bipartite optical lattice $2\times4$, and the extended Bose--Hubbard model $2\times2$; the lightly shaded horizontal bands and the corresponding connecting lines serve to distinguish the models. The red circles, blue squares, and gray diamonds represent the results of SA-$\Delta$, MPR-$\Delta$~\cite{Lin2026}, and the full cluster Gutzwiller calculation, which require 1, 4, and $N_\mu$ high-cost points on the target cluster, respectively. The speedup ratio $t_{\rm Full}/t_{\rm SA}$ is marked on the right. The three schemes use the same solver and have comparable single-point costs.}
	\label{fig:cmp-cost}
\end{figure}

\subsection{Computational cost}\label{sec:cmp-cost}
Beyond accuracy, the computational-cost advantage of SA-$\Delta$ comes primarily from the reduction in the number of high-cost solves on the target cluster. The full cluster Gutzwiller calculation requires solving $N_\mu$ high-cost critical points along the phase boundary (usually several dozen), MPR-$\Delta$ uses four training points, and SA-$\Delta$ uses only one anchor. With the same solver, Fock truncation, and convergence criterion, the average time required to solve a single critical point on the same target cluster is similar, so that the high-cost computation times of the three schemes on the target cluster scale approximately as $N_\mu:4:1$. Thus, SA-$\Delta$ reduces the high-cost computation time on the target cluster to about $1/N_\mu$ of its value for the full calculation and to about \(1/4\) of its value for MPR-$\Delta$, i.e., it accelerates the calculation by a factor of about $N_\mu$ relative to the full calculation ($31$--$50$ in the systems examined) and by a factor of about $4$ relative to MPR-$\Delta$. Figure~\ref{fig:cmp-cost} summarizes the total computation time required to solve one complete target phase boundary for the four representative systems.

The above ratio $N_\mu:4:1$ refers only to the high-cost computation time on the target cluster and does not yet include the preparatory overhead of computing the low-cost curves and performing the regression training: the preparation for MPR-$\Delta$ requires only one low-cost baseline curve and one regression training, whereas the preparation for SA-$\Delta$ requires computing two low-cost curves of smaller clusters ($J_{\rm b}$ and $J_{\rm m}$) but no regression training. However, the single-point cost of a small cluster is only about $10^{-4}$--$10^{-2}$ of that of the target cluster, so that the total time for a complete low-cost curve (two curves for SA-$\Delta$) is negligible; the training time of the MPR-$\Delta$ regression is also much shorter than that of a single critical-point calculation on the target cluster. Therefore, the ratio of the high-cost point counts on the target cluster can still serve as a reliable approximation to the ratio of the total runtimes of the schemes; the comparison of the total times on an actual platform is shown in Fig.~\ref{fig:cmp-cost}.

In all eight accuracy comparison cases of this section, SA-$\Delta$ with a single anchor outperforms MPR-$\Delta$ with its four training points. Its computational advantage comes from concentrating the high-cost solves at a single preselected anchor, while the preparatory overhead is negligible for realistic parameters, so that the time saving is determined mainly by the reduction in the number of high-cost points.

\section{Conclusions}\label{sec:conclusion}
In this paper, to address the high computational cost of phase boundaries for large clusters, we have proposed a method that combines single-anchor $\Delta$-learning (SA-$\Delta$) with the cluster Gutzwiller approximation. Its core lies in determining the shape of the correction with two low-cost curves, calibrating the amplitude with a single critical point of the target cluster, and thereby reconstructing the entire boundary with a closed-form expression. In the ten cases of the four classes of optical-lattice models, the MAPE of the reconstructed phase boundaries ranges from $0.014\%$ to $2.858\%$. In the eight cases with an MPR-$\Delta$ comparison, the errors are all lower than those of MPR-$\Delta$ with four training points. At the same time, the method reduces the number of high-cost points required per phase boundary to one---down from several dozen in the full cluster Gutzwiller calculation and from four in MPR-$\Delta$---bringing a computational speedup of about $31$--$50$ times relative to the full calculation and about $4$ times relative to MPR-$\Delta$ for the systems examined. These results show that constraining the shape of the correction with low-cost curves alone suffices both to reduce the number of target-cluster solves and to improve the reconstruction accuracy, thereby advancing the cluster Gutzwiller phase-boundary calculation from ``multipoint regression'' to ``single-point calibration plus closed-form reconstruction''. 

Building on the results for finite clusters, we further extrapolate the anchor data of clusters of several sizes to the thermodynamic limit with the aid of the finite-size scaling variable $\lambda$. The resulting boundaries agree overall with the QMC simulation results, with a visible deviation only near the tip. The results of this paper therefore support SA-$\Delta$ as an effective way to reduce the computational cost of the systems examined, and provide a viable route for efficiently predicting the phase boundaries of strongly correlated bosons in complex optical-lattice systems.

\begin{acknowledgments}
	This research is supported by the National Natural Science Foundation of China (NSFC) under Grant No.~12004005, the Scientific Research Fund for Distinguished Young Scholars of the Education Department of Anhui Province under Grant No.~2022AH020008, the Natural Science Foundation of Anhui Province under Grant No.~2008085QA26, and the open project of the State Key Laboratory of Surface Physics at Fudan University under Grant No.~KF2021\_08.
\end{acknowledgments}

\end{document}